\documentclass[twocolumn,showpacs,aps,prd,superscriptaddress,floatfix]{revtex4}
\usepackage{graphicx}
\usepackage{dcolumn}
\usepackage{bm}

\newcommand{\BABARPubYear}    {08}
\newcommand{\BABARPubNumber}  {025}
\newcommand{\SLACPubNumber} {13316}
\newcommand{\LANLNumber} {0807.3103}

\input babarsym

\newcommand{\Btokstg}{\ensuremath{\Bz\to\Kstarz\gamma}}

\newcommand{\Btokspizg}{\ensuremath{\Bz\to\KS\piz\gamma}}

\newcommand{\Bptokstpg}{\ensuremath{\Bp\to\Kstarp\gamma}}
\newcommand{\leg}{\ensuremath{L_2/L_0}}
\newcommand{\mkspiz}{\ensuremath{m(\KS \piz)}}

\newcommand{\bi}{\begin{itemize}}
\newcommand{\ei}{\end{itemize}}

\long\def\inst#1{\par\nobreak\kern 4pt\nobreak
    {\it #1}\par\vskip 10pt plus 3pt minus 3pt}

\begin{document}

\begin{flushleft}
\babar-PUB-\BABARPubYear/\BABARPubNumber \\
SLAC-PUB-\SLACPubNumber \\
arXiv:\LANLNumber\ [hep-ex]
\end{flushleft}

\title{\boldmath Measurement of Time-Dependent {\ensuremath{C\!P}\xspace} Asymmetry in $B^0 \rightarrow K^0_{\scriptscriptstyle S} \pi^0 \gamma$ Decays}

%
\author{B.~Aubert}
\author{M.~Bona}
\author{Y.~Karyotakis}
\author{J.~P.~Lees}
\author{V.~Poireau}
\author{E.~Prencipe}
\author{X.~Prudent}
\author{V.~Tisserand}
\affiliation{Laboratoire de Physique des Particules, IN2P3/CNRS et Universit\'e de Savoie, F-74941 Annecy-Le-Vieux, France }
\author{J.~Garra~Tico}
\author{E.~Grauges}
\affiliation{Universitat de Barcelona, Facultat de Fisica, Departament ECM, E-08028 Barcelona, Spain }
% ...
\collaboration{The \babar\ Collaboration}
\noaffiliation

\date{October 14, 2008}

\begin{abstract}
We measure the time-dependent {\ensuremath{C\!P}\xspace} asymmetry in
{\ensuremath{\Bz\rightarrow{\ensuremath{K^0_{\scriptscriptstyle
S}}\xspace}{\ensuremath{\pi^0}\xspace}\gamma}}\ decays for two regions
of {\ensuremath{K^0_{\scriptscriptstyle
S}}\xspace}-{\ensuremath{\pi^0}\xspace} invariant mass,
{\ensuremath{m({\ensuremath{K^0_{\scriptscriptstyle S}}\xspace}
{\ensuremath{\pi^0}\xspace})}}, using the final {\mbox{\slshape
B\kern-0.1em{\smaller A}\kern-0.1em B\kern-0.1em{\smaller A\kern-0.2em
R}}}\ data set of $467 \times 10^6$ {\ensuremath{B{\kern
0.18em\overline{\kern -0.18em B}{}\xspace}}\xspace} pairs collected at
the PEP-II {\ensuremath{e^+e^-}\xspace} collider at SLAC. We find $339
\pm 24$ {\ensuremath{\Bz\rightarrow\Kstarz\gamma}}\ candidates and measure
$S_{K^* \gamma} = -0.03 \pm 0.29 \pm 0.03$ and $C_{K^* \gamma} = -0.14
\pm 0.16 \pm 0.03$. In the range $1.1 <
{\ensuremath{m({\ensuremath{K^0_{\scriptscriptstyle S}}\xspace}
{\ensuremath{\pi^0}\xspace})}} < 1.8 \gevcc$ we find $133 \pm 20$
{\ensuremath{\Bz\rightarrow{\ensuremath{K^0_{\scriptscriptstyle
S}}\xspace}{\ensuremath{\pi^0}\xspace}\gamma}}\ candidates and measure
$S_{{\ensuremath{K^0_{\scriptscriptstyle S}}\xspace}
{\ensuremath{\pi^0}\xspace} \gamma} = -0.78 \pm 0.59 \pm 0.09$ and
$C_{{\ensuremath{K^0_{\scriptscriptstyle S}}\xspace}
{\ensuremath{\pi^0}\xspace} \gamma} = -0.36 \pm 0.33 \pm 0.04$. The
uncertainties are statistical and systematic, respectively.

\end{abstract}


\maketitle

The radiative decay \btosgam serves as a probe of physics beyond the
standard model (SM). In the SM it proceeds at leading order through a
loop diagram, making it sensitive to possible virtual contributions
from as yet undiscovered particles. Because of parity violation in the
weak interaction, the photon in \btosgam is predominantly left-handed,
while it is right-handed in the charge-conjugate decay. The photon
polarization can be determined indirectly through a measurement of
time-dependent \CP asymmetry in certain neutral decay channels. A
non-zero asymmetry $S$ due to interference between \Bz mixing and decay
diagrams is only present if both photon helicities contribute to both
\Bz and \Bzb decays~\cite{Atwood:1997zr}. $S$ is expected to be
approximately $-0.02$ in the SM~\cite{Atwood:1997zr,Ball:2006cv},
though hadronic corrections might permit it to be as large as $\pm
0.1$~\cite{Grinstein:2005nu}. Several new physics scenarios yield
large values of the asymmetry; these include left-right symmetric
models~\cite{Atwood:1997zr,Cocolicchio:1988ac} and supersymmetric
models~\cite{Chun:2000hj}. Because the SM asymmetry is small, any
significant evidence of a large asymmetry would point to a source
beyond the SM.

We present an updated measurement of the time-dependent \CP asymmetry
in \Btokspizg\ based on the final \babar\ data set of $467 \times
10^6$ $\FourS \to \BB$ decays collected at the \pep2 asymmetric-energy
\epem storage rings at SLAC. Previous measurements have
been performed by \babar~\cite{Aubert:2005bu} and
Belle~\cite{Ushiroda:2006fi}. Changes since \babar's last published
result include doubling the data set, improved track
reconstruction, better removal of background photons from \piz and
$\eta$ decays, better rejection of \Bptokstpg\
background~\cite{ref:cc}, and an improved evaluation of the systematic
uncertainties from non-signal $B$ decays. At leading order in the SM,
the \CP asymmetries of this mode do not depend on
\mkspiz~\cite{Atwood:2004jj}. However, since the aforementioned hadronic
corrections~\cite{Grinstein:2005nu} or new physics could introduce
this dependence, we split the data into two parts: the \Kstar region
with $0.8<\mkspiz<1.0\gevcc$ and the non-\Kstar region with
$1.1<\mkspiz<1.8\gevcc$.

Time-dependent \CP asymmetries are determined using the difference of
\Bz meson proper decay times $\deltat = t_{\rm sig} - t_{\rm tag}$,
where $t_{\rm sig}$ is the proper decay time of the signal \Btokspizg\
candidate ($B_{\rm sig}$) and $t_{\rm tag}$ is that of the other $B$
($B_{\rm tag}$), which is partially reconstructed and flavor-tagged
based on its daughter tracks. The \deltat\ distribution for $B_{\rm
sig}$ decaying to a \CP eigenstate is
\begin{equation}
{\cal{P}}_\pm (\Delta t) = \frac{e^{-|\Delta t|/\tau}}{4 \tau}
[1 \pm S \sin(\Delta m \Delta t) 
   \mp C \cos(\Delta m \Delta t)],
\label{eq:deltatpdf} 
\end{equation}
where the upper and lower signs correspond to $B_{\rm tag}$ having
flavor \Bz and \Bzb respectively, $\tau$ is the \Bz lifetime, and
$\Delta m$ is the \Bz-\Bzb mixing frequency. The $C$ coefficient
corresponds to the direct \CP asymmetry in decay, expected to be
smaller than 1\% in the SM~\cite{Soares:1991te}.

We evaluate our selection criteria with a detailed Monte Carlo (MC)
simulation of the \babar\ detector~\cite{Aubert:2001tu}, using the {\tt
EVTGEN} generator~\cite{Lange:2001uf} and the {\tt GEANT4}
package~\cite{Agostinelli:2002hh}.  We require photon candidates to
have energy greater than 30~\mev and the expected lateral
shower shapes in the electromagnetic calorimeter (EMC). The primary photon
from the $B$ decay 
must be isolated by more than $25\cm$ from other
charged and neutral clusters in the EMC. Primary-photon candidates
that make a \piz or $\eta$ candidate when
combined with another photon in the event are discarded based on a
likelihood formed from the diphoton mass and the energy of the second
photon. We select $\KS \to \pip
\pim$ candidates from oppositely-charged tracks for which the
probability of a geometrical vertex fit is greater than 0.1\%, the
$\pip\pim$ invariant mass is between 487 and 508~\mevcc, and the
reconstructed decay length is greater than 5 times its uncertainty. We
select $\piz
\to\gaga$ candidates with invariant mass between 115 and 155~\mevcc
and energy greater than 590~\mev in the laboratory frame. For
candidates in the \Kstar region we require
$|\cos{\theta_{\Kstar}}| < 0.9$, where $\theta_{\Kstar}$ is the angle
between the \KS and primary photon direction in the \Kstar rest frame.

To identify signal decays we use the energy-substituted mass $\mes =
\sqrt{ (s/2c^2 + {\bf p}_0 \cdot {\bf p}_B)^2/E^2_0 - |{\bf
p}_B|^2/c^2}$ and the energy difference $\DeltaE = E^*_B -
\sqrt{s}/2$, where $(E_0/c,{\bf p}_0)$ and $(E_B/c, {\bf p}_B)$ are
the four-momenta of the initial \epem system and the $B$ candidate,
respectively, $\sqrt{s}$ is the center-of-mass (CM) energy, and the
asterisk denotes the CM frame. The distributions of signal events show
a peak in these variables. We require $5.2 < \mes < 5.3\gevcc$ and
$|\DeltaE| < 250\mev$. To reduce $\Bp \to \Kstarp \gamma$ background,
we reconstruct $B^+
\to\Kstarp(\KS \pip)\gamma$ candidates subject to the same
requirements as \Bz candidates, and veto events for which $\mes(\Bp) >
5.27\gevcc$ and $0.8<m(\KS\pip)<1.0\gevcc$. To discriminate $B$ decays
from continuum $\epem \to \qqbar$ ($q = u,d,s,c$) background, we
require $|\cos{\theta^*_B}| < 0.9$, where $\theta^*_B$ is the CM angle
between the $B$ candidate and the $e^-$ beam direction. We require the
ratio of event-shape moments \leg\ to be less than 0.55, where $L_i = \sum_j
|p^*_j||\cos{\theta^*_j}|^i$, $p^*_j$ is the CM momentum of each
particle $j$ not used to reconstruct the $B$ candidate, and
$\theta^*_j$ is the CM angle between $p^*_j$ and the thrust axis of
the reconstructed $B$ candidate. After all selection criteria have
been applied we find 10587 candidate events, 16\% of which have more
than one signal \Bz candidate. In these cases we select the one with
\piz mass closest to its nominal value~\cite{Yao:2006px}, and if there
is still an ambiguity, we select the one with the \KS mass closest to
its nominal value.  We find an overall selection efficiency of 16\%.

For each reconstructed signal candidate we use the remaining tracks in
the event to determine the decay vertex position and flavor of $B_{\rm
tag}$. The latter is determined by a neural network based on kinematic
and particle identification information, the performance of which is
evaluated using a sample of fully-reconstructed, self-tagging hadronic
\Bz decays ($B_{\rm flav}$ sample)~\cite{Aubert:2007hm}.

We determine the proper time difference between $B_{\rm sig}$ and
$B_{\rm tag}$ from the spatial separation between their decay vertices
in the same way as our previous analysis and a similar \babar\ study
of $\Bz\to\KS\piz$~\cite{Aubert:2004xf}. Because both the transverse
flight length of the \Bz mesons and the transverse size of the
interaction region are small compared to the \Bz flight length along
the boost direction, we are able to determine a decay vertex from the
intersection of the \KS trajectory with the interaction region. We
further improve the \deltat resolution by 11\% over what is obtained
using information from the interaction region alone by refitting the
$\FourS\to\Bz\Bzb$ system with the constraint that the average sum of
decay times $(t_{\rm sig} + t_{\rm tag})$ be equal to twice the B
lifetime with an uncertainty of $\sqrt{2}\tau_B$. Using MC simulation
data we verify that this procedure gives an unbiased estimate of
\deltat. We define events as having good \deltat quality if each pion
daughter of the \KS creates at least 2 hits in the silicon vertex
tracker (SVT), and if the \deltat uncertainty $\sigma_{\deltat} <
2.5$~ps and $|\deltat| < 20$~ps. About 70\% of signal and background
events pass these requirements. We split our data set and fitting
procedure based on the \deltat quality such that flavor-tagged events
with poor \deltat information do not contribute to the measurement of
$S$, but do contribute to the measurement of $C$, which can be
determined solely through tagging.

We extract signal yields and \CP asymmetries using an unbinned maximum
likelihood fit to \mes, \DeltaE, \leg, tag flavor, \deltat,
$\sigma_{\deltat}$, and, in the \Kstar region, \mkspiz. Continuum and
\BB backgrounds are also modeled in the fit. We construct
the likelihood function for each contribution as the
product of one-dimensional probability density functions (PDFs). The
signal PDFs in \mes and \DeltaE are parametrized using the function
\begin{equation}
f(x) = \exp\left[\frac{-(x-\mu)^2}{2\sigma^2 + \alpha(x-\mu)^2}\right],
\end{equation}
where $\mu$ is the mean, $\sigma$ the core width, and $\alpha$ a tail
parameter. The latter two parameters are allowed to be different on
either side of the peak. The signal \mkspiz\ shape is a relativistic
Breit-Wigner, as nonresonant contributions in the $K^*$ region are
negligible~\cite{Aubert:2004te}. For continuum \mes we use an
ARGUS~\cite{Albrecht:1990cs} function, while for \DeltaE we use an
exponential shape. The continuum and \BB shape in \mkspiz\ is a
Breit-Wigner on top of a linear background. We parameterize the \BB
\mes shape as the sum of an ARGUS function and a Gaussian with
different widths below and above the peak, and the \DeltaE shape as an
exponential. The \leg\ shapes are binned PDFs, and in those the signal
and \BB components share the same parameters. All signal and \BB PDF
parameters are determined using simulated events, except for the
flavor tag efficiencies, mistag probabilities, and \deltat resolution
function parameters, which are determined from the $B_{\rm flav}$
sample. The large number of continuum background events in the fit
determine the continuum PDF parameters.

We obtain the \deltat PDF for signal events and \BB background from
Eq.~\ref{eq:deltatpdf}, accounting for the mistag probability and
convolving with the \deltat resolution function, which is the sum of
three Gaussian distributions~\cite{Aubert:2007hm}. The effective \CP
asymmetries for the \BB background, $S^{\rm bkg}_{\BB}$ and $C^{\rm
bkg}_{\BB}$, are fixed to zero in the fit, and we account for a
possible deviation from zero in the systematic uncertainty. We verify
in simulation that the parameters of the resolution function for
signal events are compatible with those obtained from the $B_{\rm
flav}$ sample. Therefore we use the $B_{\rm flav}$ parameters for
better precision. We fit the continuum MC \deltat distribution and
find that it is well-modeled by a prompt decay distribution consisting
only of the \deltat resolution function shape. The parameters of the
continuum \deltat PDF are determined in the fit to data.

\begin{figure}
\begin{center}
    \includegraphics[width=0.45\linewidth]{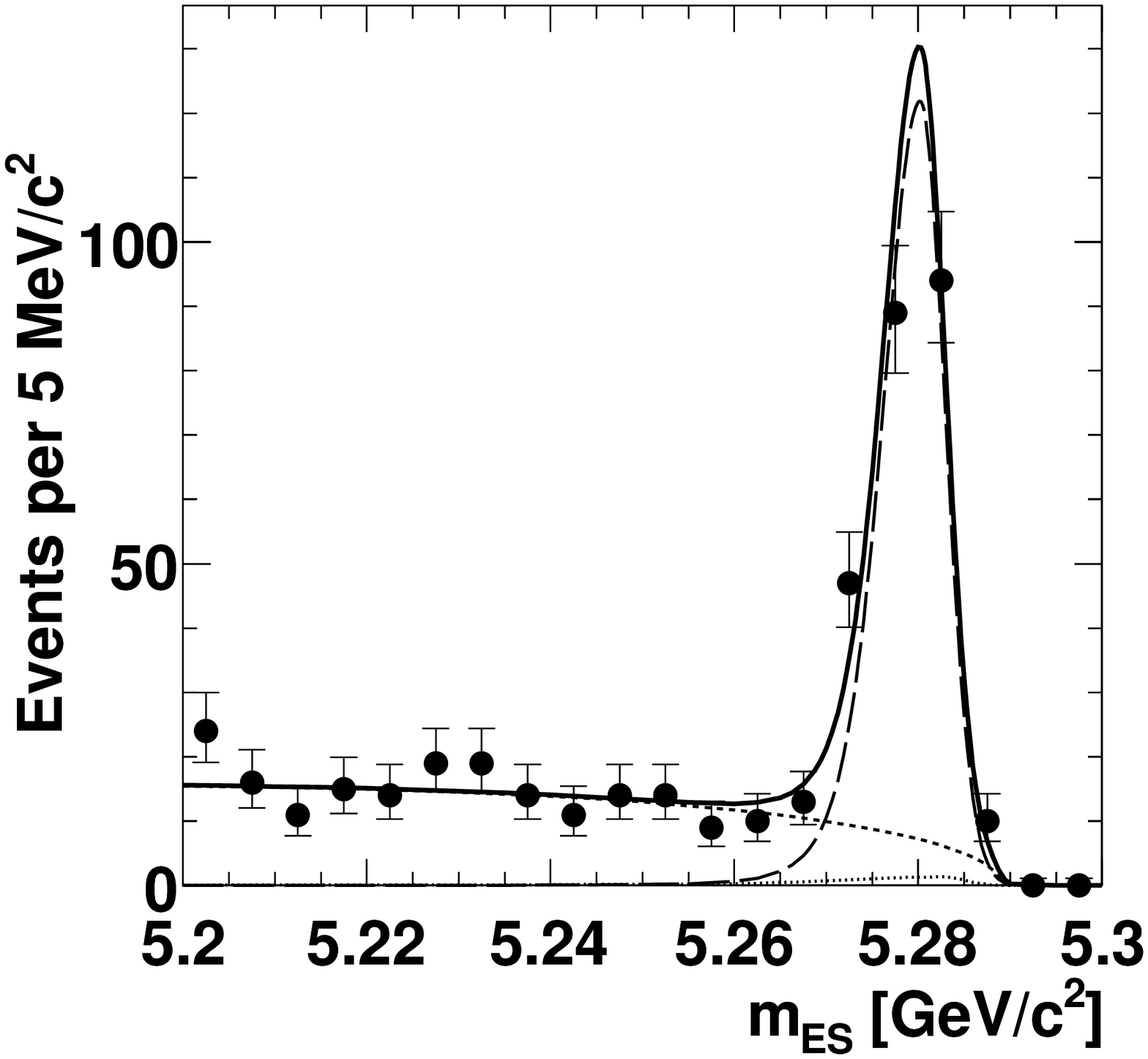}
    \includegraphics[width=0.45\linewidth]{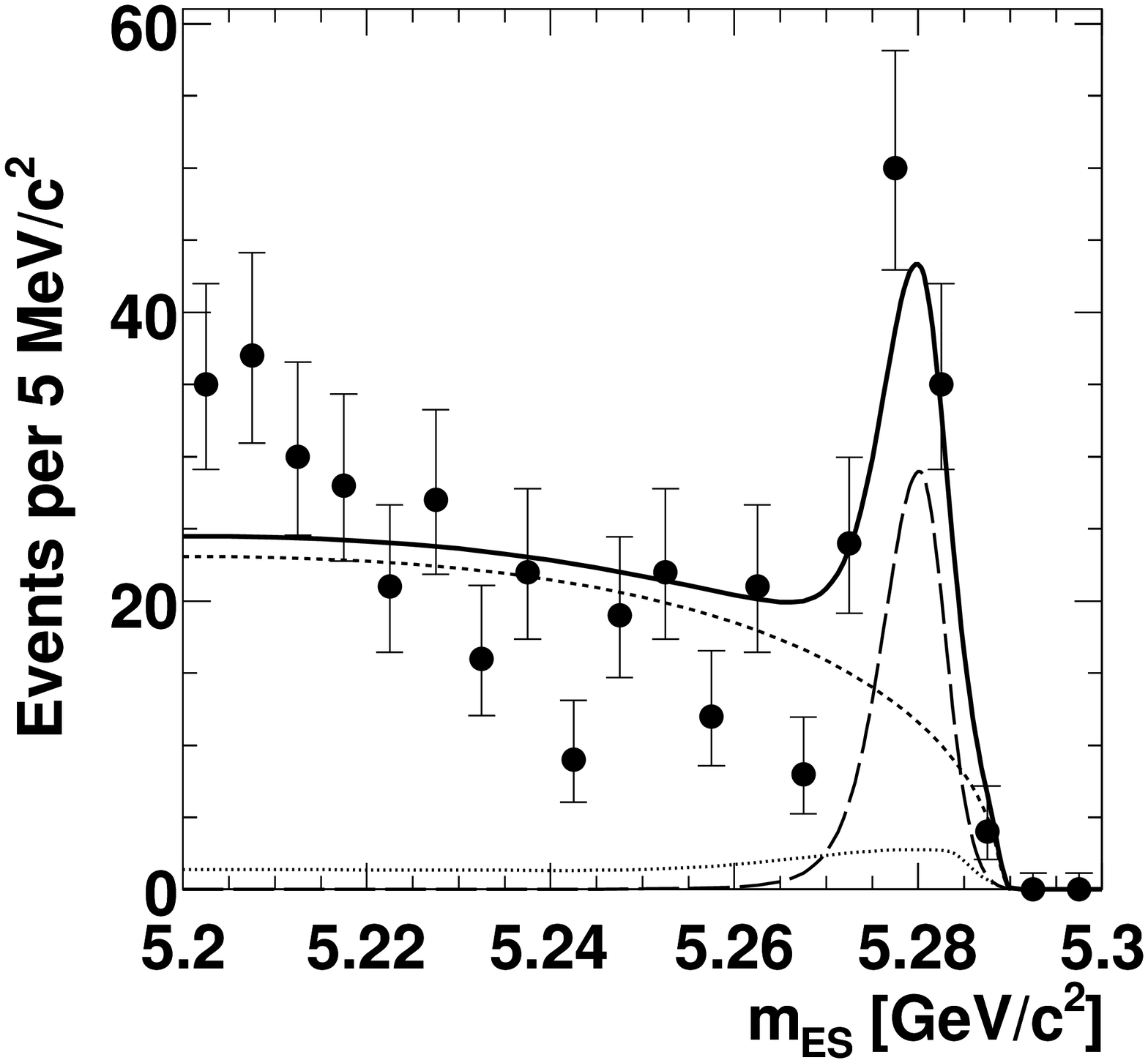}
    \includegraphics[width=0.45\linewidth]{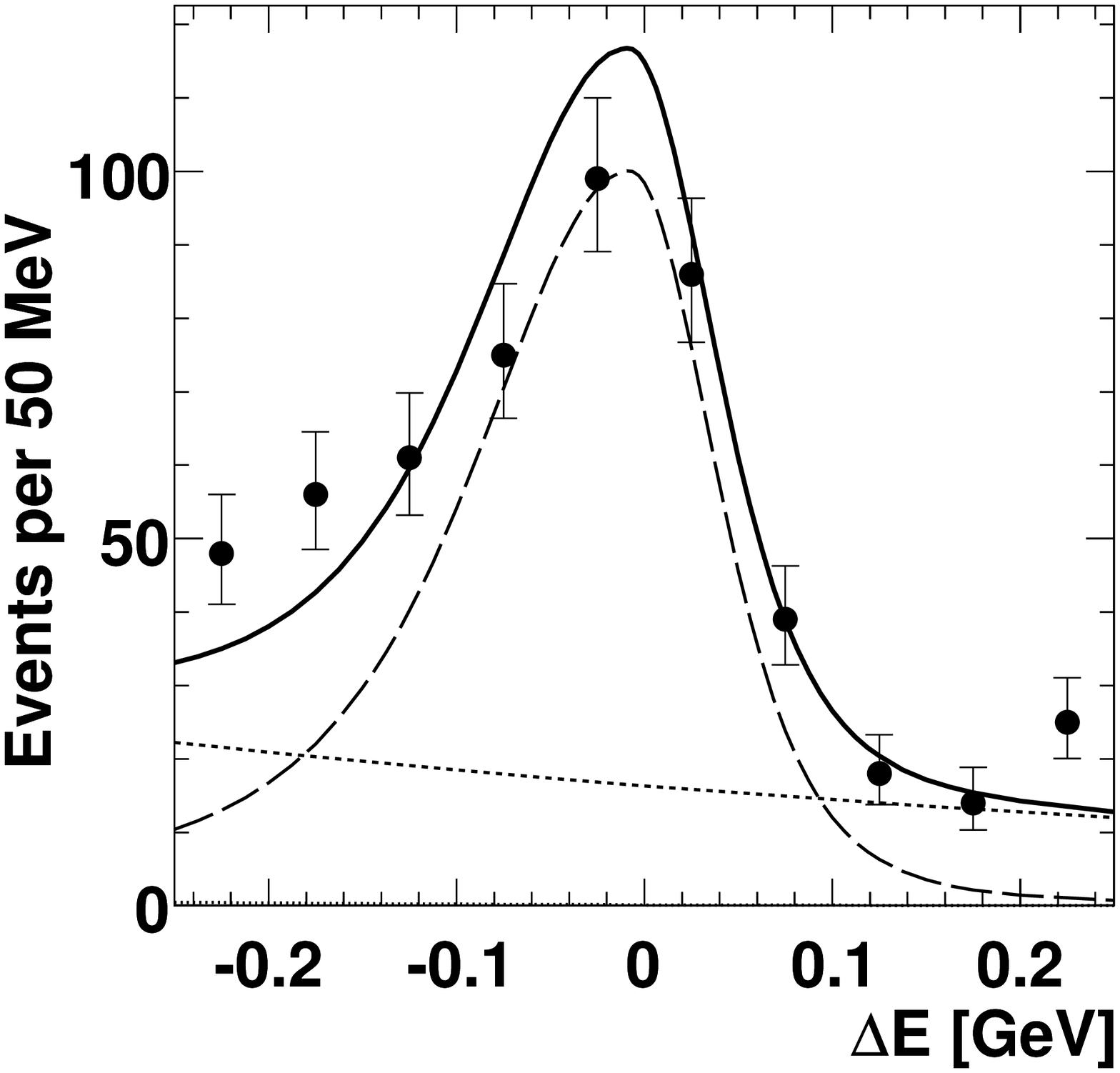}
    \includegraphics[width=0.45\linewidth]{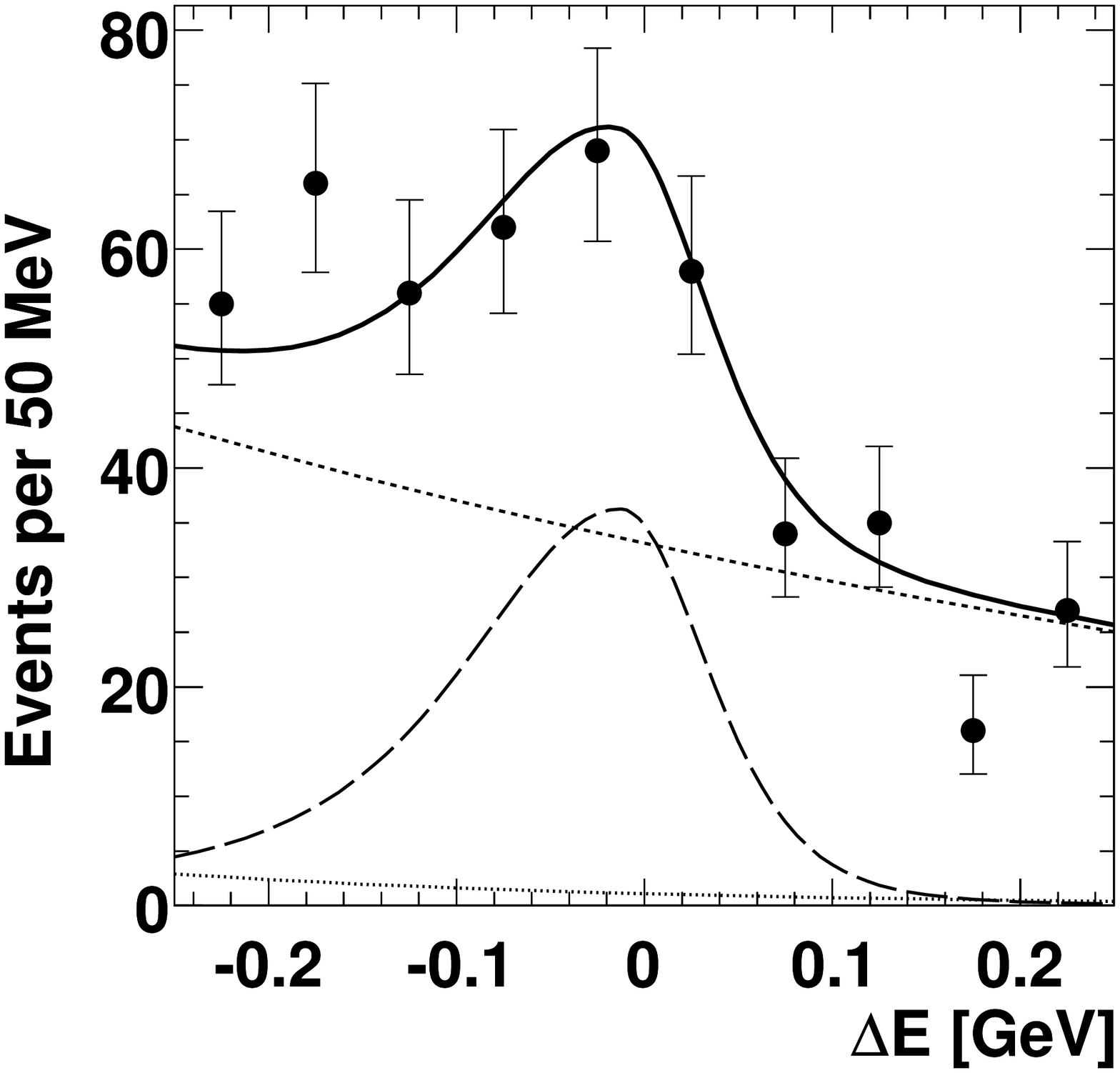}
\end{center}
  \caption{ 
Signal-enhanced distributions for \mes (top) and \DeltaE{} (bottom) for the \Kstar region (left) and the non-\Kstar region (right). We show the fit result (solid line) and PDFs for signal (long dashed), continuum (short dashed), and \BB (dotted).}
  \label{fig:fitMesDe}
\end{figure}

In the fit to the \Btokstg\ candidate sample of 3884 events we find
$339 \pm 24\ \stat$ signal events, $S_{\Kstar \gamma} = -0.03 \pm
0.29\ \stat \pm 0.03\ \syst$ and $C_{\Kstar\gamma} = -0.14 \pm 0.16\ \stat
\pm 0.03\ \syst$. We also find $19\pm 27\ \stat$ \BB background
events. In the range $1.1 < \mkspiz < 1.8 \gevcc$ with 6703 events we
measure $133 \pm 20\ \stat$ signal events, $S_{\KS \piz \gamma} = -0.78
\pm 0.59\ \stat \pm 0.09 \ \syst$ and $C_{\KS \piz \gamma} = -0.36 \pm
0.33\ \stat \pm 0.04\ \syst$. We find $167\pm 49\ \stat$
\BB background events in this sample. 
The linear correlation coefficient between $S_{\Kstar \gamma}$ and
$C_{\Kstar\gamma}$ is $+0.050$, while for $S_{\KS\piz \gamma}$ and
$C_{\KS\piz\gamma}$ it is $+0.015$. Figure~\ref{fig:fitMesDe} shows
signal-enhanced distributions for \mes\ and \DeltaE\, created by
cutting on the likelihood of the unplotted fit variables. 

We perform a cross-check because of the discrepancy between the
projection of the fit model and the data in the non-$K^*$ region at
low \mes. A fit of the data sample with $\mes > 5.22\gevcc$ shows that
the observed changes in $S$ and $C$ are consistent with statistical
fluctuations, so the signal is not significantly
affected. Additionally, we verified that the slope of the \mes\
background shape is not correlated with the other fit variables. MC
simulations of common \BB backgrounds, including the final states $\KS
\pi \pi \gamma$ and $\KS \pi \pi \piz$, do not show any rising
structure at low \mes.

\begin{figure}
  \begin{center}
  \includegraphics[width=0.45\linewidth]{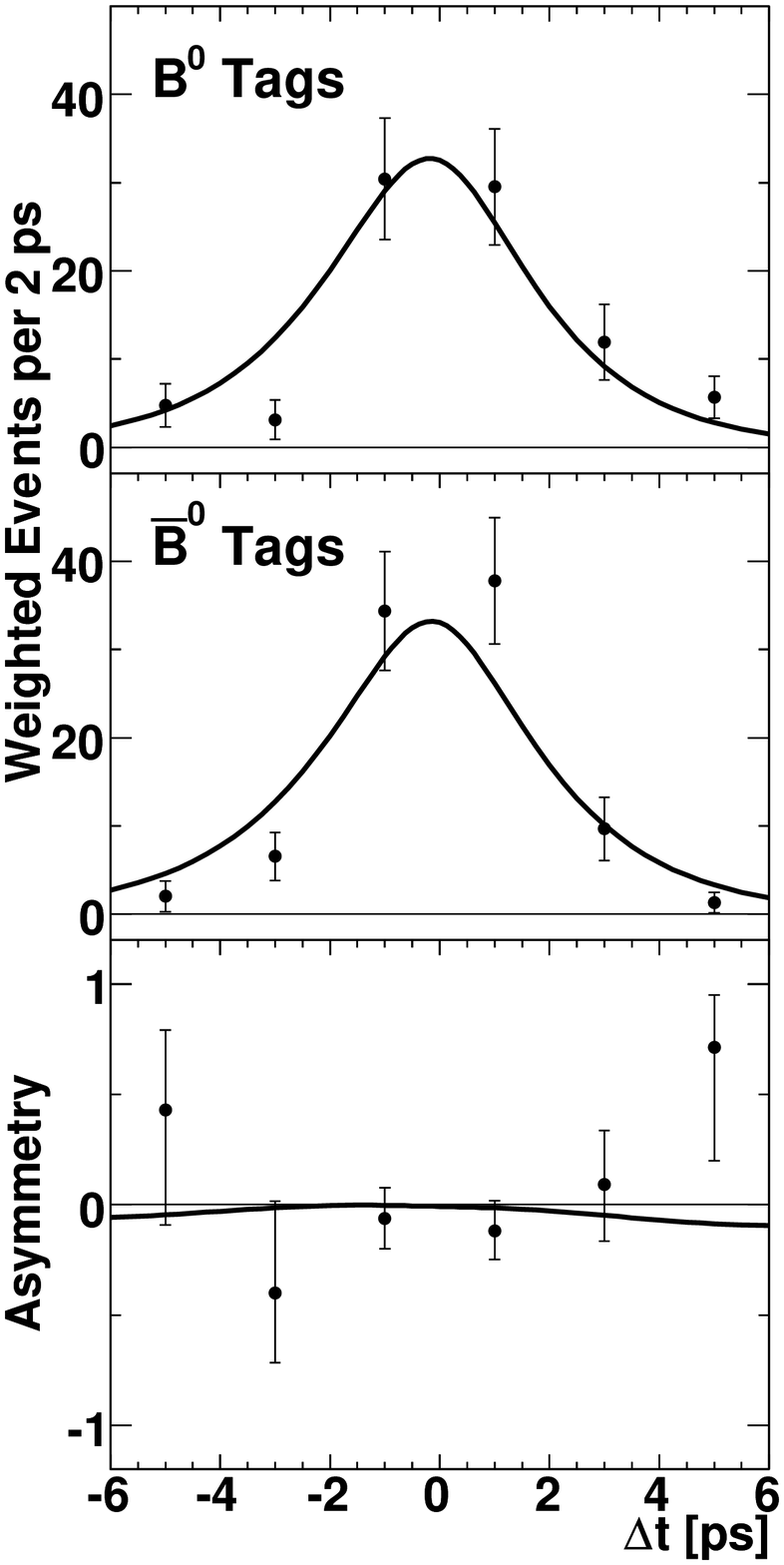}
  \includegraphics[width=0.45\linewidth]{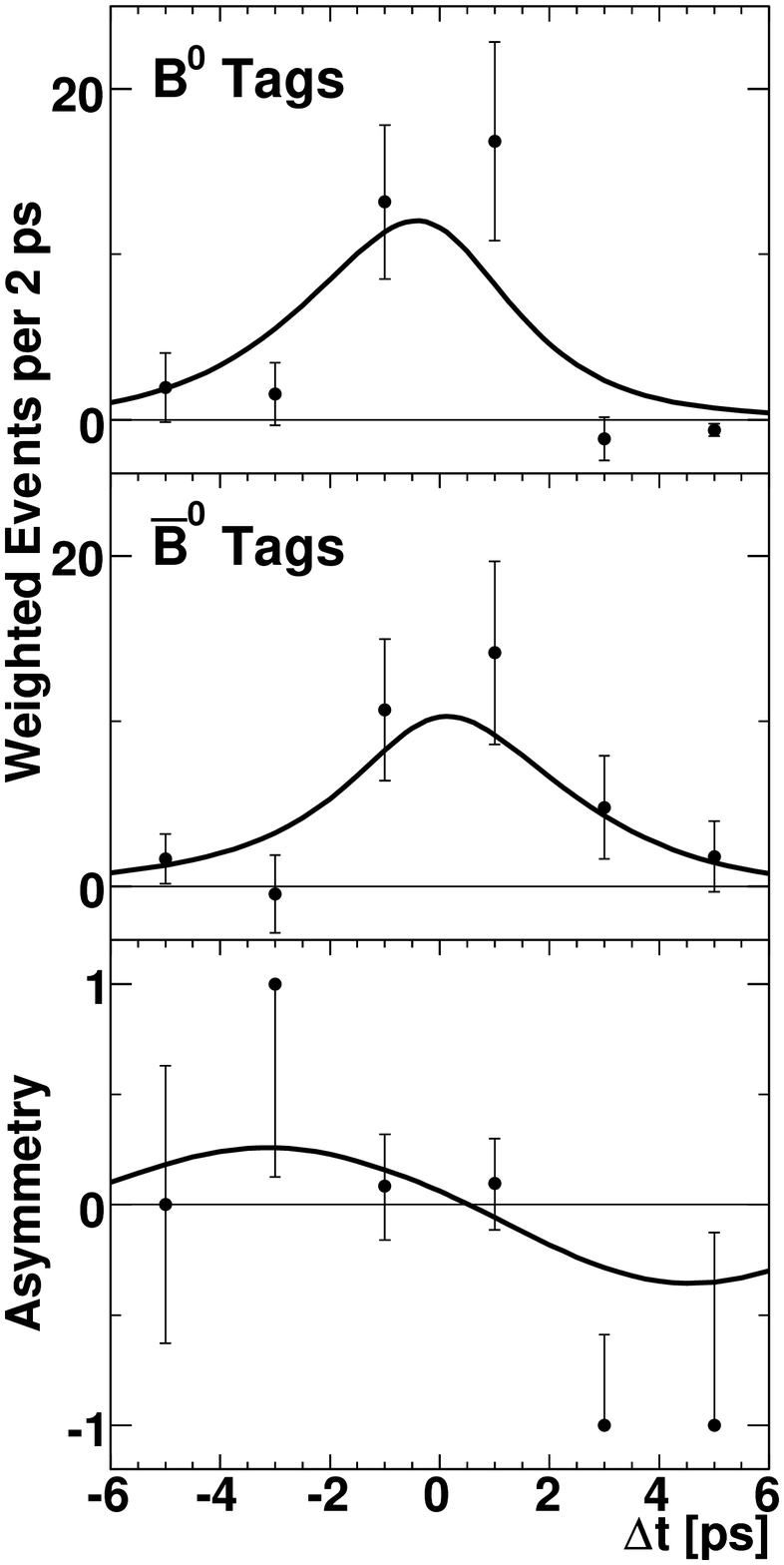} \end{center}
  \caption{ 
  sPlot (see text) of $\deltat$ in the \Kstar region (left) and the
  non-\Kstar region (right), with $B_{\rm tag}$ tagged as $\Bz$ (top)
  or $\Bzb$ (center), and the asymmetry (bottom). The curves are the
  signal PDFs.}  \label{fig:fitDt}
\end{figure}

\begin{figure}
  \centerline{
  \includegraphics[width=0.60\linewidth]{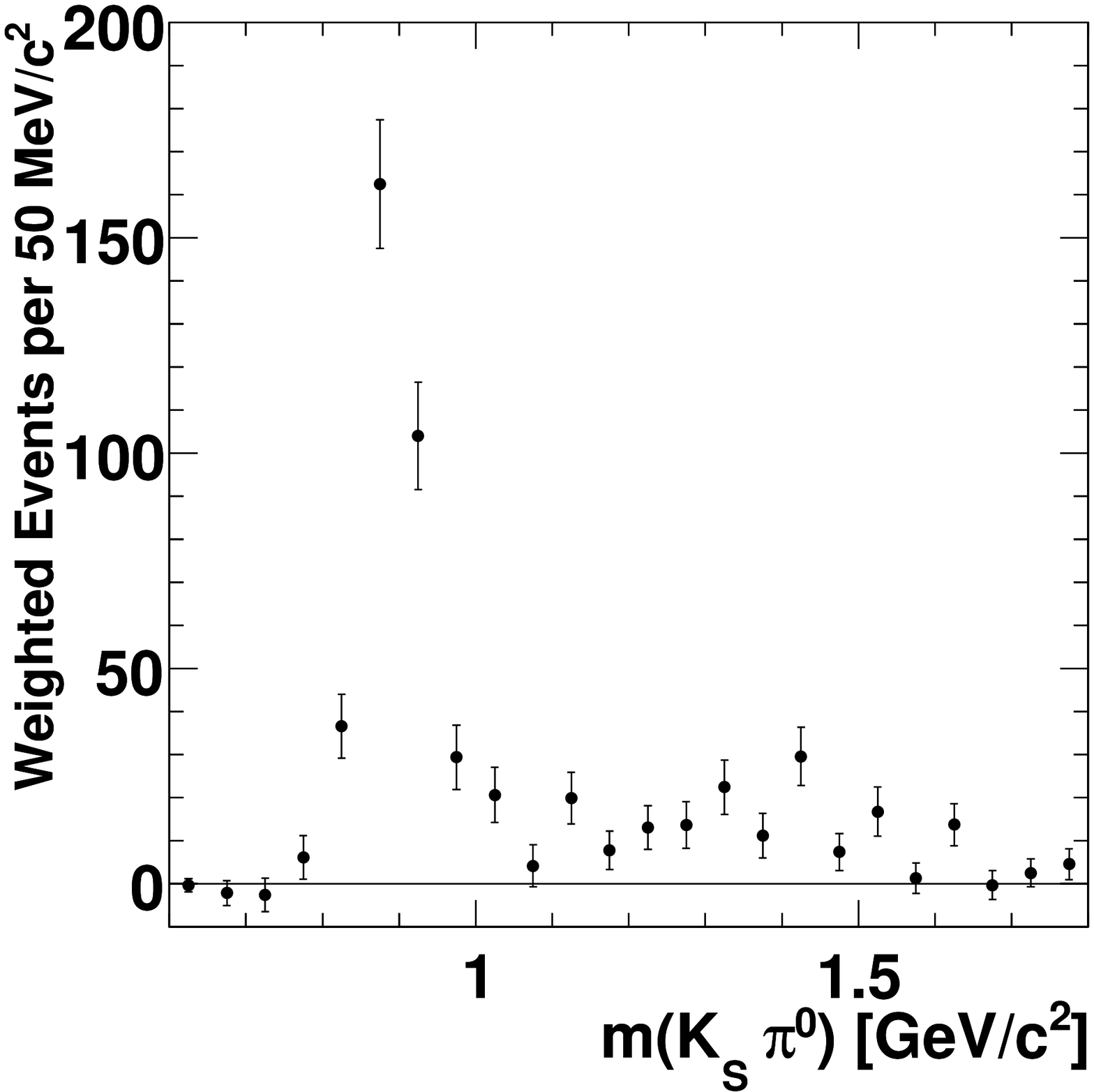}}
  \caption{sPlot (see text) of $\mkspiz$.}
\label{fig:mkspizsplot}
\end{figure}

Figure~\ref{fig:fitDt} shows the background-subtracted distributions
of \deltat in the \Kstar region, obtained with the sPlot event
weighting technique~\cite{Pivk:2004ty}. We show an sPlot of the
\mkspiz\ spectrum in Fig.~\ref{fig:mkspizsplot}.

Using an ensemble of simulated experiments generated from the fitted
likelihood function we find no bias in the \Kstar region and a spread
in $S$ and $C$ consistent with the statistical uncertainties. In the
non-\Kstar region we find a bias of $-0.06 \pm 0.03$ on $S$ and a
spread in $C$ larger than the statistical uncertainty reported by the
fit. These effects are due to a measurement that is close to the
physical boundary of $S^2 + C^2 \leq 1$, and they disappear if we
generate the ensemble with $S_{\KS\piz\gamma} = C_{\KS \piz \gamma} =
0$. The bias on $S_{\KS\piz\gamma}$ is evaluated with several
ensembles of simulated events in which the generated
$S_{\KS\piz\gamma}$ is varied. For the statistical uncertainty on $C$
we take the ensemble's root-mean-square width of $0.33$ instead of the
0.29 uncertainty determined by the fit to data.

Systematic uncertainties associated with our knowledge of the beam
spot position and possible SVT misalignment are determined by varying
the beam spot and SVT alignment parameters in MC. We bound the effects
of uncertainties in the \deltat resolution function due to the
vertexing method with a study from \babar's $\Bz\to\KS\piz$
analysis~\cite{Aubert:2007mgb}. Resolution function differences
between data and MC in control samples of $\Bz\to\jpsi\KS$ decays, in
which the \jpsi vertex information is ignored, lead to differences in
$S$ and $C$ that we take as systematic uncertainties. Uncertainties
from doubly-Cabibbo-suppressed (DCS) decays of the $B_{\rm tag}$ are
included as in Ref.~\cite{Aubert:2007hm}.

We evaluate uncertainties due to the vertex reconstruction procedure
and possible correlations among the observables with an ensemble of
simulated experiments created by generating background events from the
PDFs and embedding signal events from the full MC simulation. No
significant bias is observed in the \Kstar region, and we bound
uncertainties by the precision with which the potential bias is
measured. In the non-\Kstar region, no bias is observed in the
signal MC.

Uncertainties due to limited knowledge of the fixed parameters in the
fit are evaluated by varying them within their uncertainties. We
evaluate differences between data and MC in the signal shape by fixing
the background parameters to those determined in the fit to data and
floating the signal parameters separately for each observable.

We evaluate the effect of $S^{\rm bkg}_{\BB}$ and $C^{\rm bkg}_{\BB}$
by varying them over a range determined by the composition of the \BB
background samples and \CP asymmetry measurements in the PDG
listings. The systematic uncertainties are summarized in
Table~\ref{tab:systematics}.

\begin{table}
  \caption{Summary of systematic uncertainties. \label{tab:systematics} }
\begin{ruledtabular}
    \begin{tabular}{lcccc}
        & \multicolumn{2}{c}{\hspace*{1.5em}\Kstar Region\hspace*{1.5em}} & \multicolumn{2}{c}{non-\Kstar Region} \\
\hline
        Source & $\Delta S$ & $\Delta C$ & $\Delta S$ & $\Delta C$ \\
      \hline
        Beamspot                          & 0.007 & 0.002 & 0.007 & 0.002 \\
        SVT Alignment                     & 0.010 & 0.010 & 0.010 & 0.010 \\
        Resolution Function               & 0.011 & 0.018 & 0.011 & 0.018 \\
        Bias Uncertainty                  & 0.015 & 0.009 & 0.028 & 0.016 \\
        PDF Uncertainty                   & 0.015 & 0.013 & 0.060 & 0.019 \\
        $S^{\rm bkg}_{\BB}$ and $C^{\rm bkg}_{\BB}$ & 0.008 & 0.002 & 0.060 & 0.018 \\
        DCS $B_{\rm tag}$ Decays          & 0.001 & 0.015 & 0.001 & 0.015 \\
        \hline			       
        Total                             & 0.028 & 0.030 & 0.091 & 0.040 \\
\end{tabular}
\end{ruledtabular}
\end{table}

In summary, we have measured the time-dependent \CP asymmetry in
\Btokspizg\ decays using the full \babar\ data set recorded at the
\FourS resonance. We find
\begin{eqnarray*}
S_{K^* \gamma} &=& -0.03 \pm 0.29\ \stat \pm 0.03\ \syst,\\
C_{K^* \gamma} &=& -0.14 \pm 0.16\ \stat \pm 0.03\ \syst,\\
S_{\KS \piz \gamma} &=& -0.78 \pm 0.59\ \stat \pm 0.09\ \syst,\\
C_{\KS \piz \gamma} &=& -0.36 \pm 0.33\ \stat \pm 0.04\ \syst.
\end{eqnarray*}
The measurement in each \mkspiz\ region is consistent within
uncertainties with the predictions of the standard model.

We are grateful for the excellent luminosity and machine conditions
provided by our \pep2\ colleagues, 
and for the substantial dedicated effort from
the computing organizations that support \babar.
The collaborating institutions wish to thank 
SLAC for its support and kind hospitality. 
This work is supported by
DOE
and NSF (USA),
NSERC (Canada),
CEA and
CNRS-IN2P3
(France),
BMBF and DFG
(Germany),
INFN (Italy),
FOM (The Netherlands),
NFR (Norway),
MES (Russia),
MEC (Spain), and
STFC (United Kingdom). 
Individuals have received support from the
Marie Curie EIF (European Union) and
the A.~P.~Sloan Foundation.


\begin{thebibliography}{99}
\bibitem{ref:cc}
Unless explicitly stated, charge conjugate decay modes are included
implicitly throughout this paper.
\end{thebibliography}
\end{document}